\documentclass[conference]{IEEEtran}
\ifCLASSINFOpdf
  \usepackage[pdftex]{graphicx}
  \DeclareGraphicsExtensions{.pdf,.jpeg,.png}
\else
\fi
\usepackage[tight,footnotesize]{subfigure}
\usepackage{url}
\usepackage{pifont}
\usepackage{tabulary}

\begin{document}
%
\title{Potential mass surveillance and privacy violations in proximity-based social applications}

\author{\IEEEauthorblockN{Silvia Puglisi, David Rebollo-Monedero and Jordi Forn\'e}
\IEEEauthorblockA{Department of Telematics Engineering, \\ 
Universitat Polit\`ecnica de Catalunya (UPC)\\
C.\ Jordi Girona 1-3, 08034 Barcelona, Spain\\
silvia.puglisi@upc.edu\\
david.rebollo@entel.upc.edu\\
jforne@entel.upc.edu
}}

\maketitle

\begin{abstract}
Proximity-based social applications let users interact with people that are currently close to them, by revealing some information about their preferences and whereabouts. This information is acquired through passive geo-localisation and used to build a sense of serendipitous discovery of people, places and interests.
Unfortunately, while this class of applications opens different interactions possibilities for people in urban settings, obtaining access to certain identity information could lead a possible privacy attacker to identify and follow a user in their movements in a specific period of time. The same information shared through the platform could also help an attacker to link the victim's online profiles to physical identities.
We analyse a set of popular dating application that shares users relative distances within a certain radius and show how, by using the information shared on these platforms, it is possible to formalise a multilateration attack, able to identify the user actual position. The same attack can also be used to follow a user in all their movements within a certain period of time, therefore identifying their habits and Points of Interest across the city. Furthermore we introduce a social attack which uses common Facebook likes to profile a person and finally identify their real identity.

\end{abstract}


%
\IEEEpeerreviewmaketitle

\section{Introduction}
The communication possibilities opened by online services are almost endless. Social media allow people everyday to know more about themselves, their friends and their surrounding. To use such services, users grant them a certain level of access to their private data. This data include details about their identity, their whereabouts and in some situations even the company they work for. This level of access is obtained leveraging on third parties, like Facebook or Google, which offer login technologies, allowing the application to identify the user and receive precise information about them.
Once the user grant access to their data, the application stores it and assumes control over how it is further shared. The user will never be notified again on who is accessing their data, nor if these are transferred to third parties. 
We believe this can expose users of such services to privacy attacks, while in addition preventing them to retain direct control over their data and who has access to it over time.
This aspect of privacy protection is particularly relevant since usually the right to privacy is interpreted as the user's right to prevent information disclosure. Online services use this interpretation to ask the user to access certain information, yet no concrete information is passed on how the data will be used or stored. Furthermore, these services are often designed as mobile applications where all the devices installing the app communicate with a centralised server and constantly exchange users' information, eventually allowing for unknown third parties, or potential attackers, to fetch and store this data. In addition, this information is often shared with insecure communication through the HTTP protocol, making it possible for a malicious entity to intercept these communication and steal user data.

\subsection{Contribution}

We have observed how proximity-based social applications have access to certain identity information that could lead a possible privacy attacker to easily identify users and link their online profiles to physical identities. In our study we analyse a set of popular dating application, which are built on the assumption that users can preserve a certain level of privacy by only sharing their relative distance with other users on the platform. Furthermore the user also shares Facebook likes or common categories of interests. \\
\\
\noindent
These application are built on the notion of serendipitous discovery of people, places and interests around the user's surrounding. \\
\noindent
We consider these applications an example of how many privacy violation users are subjected to without being aware of it. Furthermore, this scenario offers a playground to prove how little details about the user's whereabouts and personal sensitive information are needed to track the user and discover their real identities.  For example we prove how the user's relative distance or their first name and what common interest their share on Facebook, can allow an attacker to follow them along the day and across their movements, or even profile their full interests and discover personal details about them. \\
\\
\noindent
The main contributions of this paper are the following.

\begin{enumerate}
 \item We classify privacy threats in these applications following the categorisation of Daniel J. Solove in ~\cite{solove2006taxonomy}, which presenting a taxonomy to understand privacy violations and to identify privacy problems in a comprehensive and concrete manner.
\\ 
 \item We formalise an attack showing how proximity based social application are inherently insecure. Our attack retrieves information about nearby users, stores certain information about them, and subsequently uses these to retrieve their updated profiles at regular intervals. Our attacker agent is also able to change their relative position at will and therefore can easily perform a multilateration attack and identify the victim position with a arbitrary precision. Furthermore the attacker can keep following the user, eventually categorising their interests, movements and even identifying their Points of Interest (POI) around the city. 
\\
 \item We build a Social Graph attack using Facebook likes to know the victim interests. The applications examined, in fact, allow the attacker to know ~\emph{what they have in common} with the victim and use the known expressed interests to identify the user's Facebook profile through their Graph Search while also profiling individuals nearby.
\end{enumerate}

\section{Background}

Online communications in general and social media in particular, are increasingly opening up new possibilities for users to share and interact with people and content online. At the same time, social networking services collect and share valuable information regarding locations, browsing habits, communication records, health information, financial information, and general preferences regarding user online and offline activities. This level of access is often directly granted from the user of such services, although the privacy and sensitiveness of the information becoming accessible to third parties can be easily overlooked.\\
\\
\noindent     
Furthermore, social network are no longer a novelty and user have become used to share their information with both social relationships as well as third party applications. \\
\noindent
Leveraging on this perception of social media by Internet users, another class of applications is being developed based on the concept of ~\emph{serendipitous} discoveries. The idea of ~\emph{serendipity} in mobile applications wants the user to accidentally discover people, places and/or interests around them, by using passive geo-localisation and recommendation systems. Passive geo-localisation is a mechanism using the ability of mobile devices to know the user's position without having to constantly asking for it. Technologies that provide this capability are GPS, wireless and mobile networks, iBeacon and so on.\\
\\
\noindent
To present the user with a tailored and seamless experience, serendipity applications need to learn the user's preferences and interests. This is usually accomplished by connecting several of the user's identities on other social networks. A typical example is asking the user to register onto an application through their Facebook, Twitter, or Google+ accounts. This technique usually consists in a variant of the OAuth2.0 protocol used to confirm a person's identity and to control what data they will share with the application requesting login. \\
\\
\noindent
We have specifically analysed Facebook login since it was the common login mechanism offered in all applications examined, although the same functionalities apply for other third party login mechanisms. Facebook login provides both authentication and authorisation. The mechanism is used on the web as well as on iOS and Android, although on those platforms the primary mechanism uses the native Facebook application instead of the web API.\\
\\
\noindent
When an application is connected to the user's Facebook profile using Facebook Login, it can always access their \emph{public profile} information. Facebook consider this information public and will not apply any restriction on it. Information that is shared with the public profile vary from user to user and depends on their privacy settings. By default the Facebook public profile includes some basic attributes about the person such as the user's age range, language and country, but also the name, gender, username and user ID (account number), profile picture, cover photo and networks.\\
\\
\noindent
An application may also ask for more information about the user. These can include the list of friends using the app, their email, the events that they are attending, their hometown or the things they have liked. This information can be obtained by requesting for optional permissions, which are asked for during login process. Apps can also ask for additional permissions later, after a person has logged in.\\
\\
\noindent
The information obtained from Facebook is often displayed on the application platform or used to match people with similar interests, thus giving away more hints about an individual real identity. For example a user \emph{swiping} through other people on \emph{Tinder}~\cite{tinder} will know if they have liked similar pages on Facebook.
These hints or traces can be used to further identify that individual on other platforms. In fact, this information crossed with the city the user lives in, the user's photo, and their first name could already be enough to identify their Facebook profile. \\
\\
\noindent
The attacker could hence use what they know about the user to identify a number of profiles of people living in a certain city. A query of the form \emph{people named John who live in Barcelona and like surfing and volleyball} could be used to restrict the attacker's search space to a smaller number of profiles. Finally, since these applications tend to fetch  the profile photo directly from Facebook, the actual user profile can be identified by matching the two profile pictures.\\
\noindent
Notice that while some queries might seem very generic, some others might already restrict significantly the set of targeted profiles. It is particularly concerning in fact that these applications might be used to target specific individuals with the objective to reach confidential information about their actual job or company they work for, as reported recently by IBM in a report about security of dating apps~\cite{ibm2015report}.\\
\\
\noindent
The ubiquitous streams of data that users create while they use different application can be seen as a network of interconnected data snippets. Information shared on the web can be linked together so that it is possible to construct semantic connections between user's activity data. A possible attacker could therefore try to link data between different source of information to identify and target users both online and offline. Users become more frequently exposed to social engineering attacks that can now leverage on facts gathered online about their personal offline lives. \\
\noindent

\subsection{State of the art}

Users of Social networks should be particularly careful with the information they share on Social Network, as it has been show how leaking bits of personal information  can be used for concrete privacy attacks. For example, physical identification and password recovery attacks can be based on the knowledge of personal information or the use of a known secret~\cite{irani-et-al}. It has been shown how the attribute set {birth-date, gender, zip code} poses concrete risks of individual identification~\cite{sweeney}, leading to details that can be used to identify physical persons or to infer answers to password­ recovery questions. \\
\\
\noindent
Another important aspect to consider is that the average online user joins different social networks with the objective to enjoy distinct services and features. On each service or application an identity gets created, containing personal details, preferences, generated content and a network of relationships. The set of attributes used to describe these identities is often unique to the user. In addition application or services sometimes require the disclosure of different personal information, such as email or full name, to create a profile. Users possessing different identities on different services, often use those to verify another identity on a particular application, i.e. a user will use their Facebook and LinkedIN profile to verify their account on the third service~\cite{paridhi-et-al}. A set of information required by one service could, in fact, add credibility to the information the user has provided for a second application, by demonstrating that certain personal details overlap, and by adding other information, like, for example, a set of shared social relationships. \\
\\
\noindent
Users online footprints could therefore be reconstructed by combining the publicly available information provided to different services~\cite{irani-et-al-2}~\cite{goga-et-al}. A possible attacker could start by identifying common pseudonym, i.e. a username that users often use across different social networks, then goes on measuring how many possible profiles it can find across different services. 
Therefore, a user's activity on one site can implicitly reveal their identity onto another site, also investigating how locations attached to posts could be used uniquely to identify a profile among a certain number of similar candidates.
\\
\noindent
The analysis of publicly available attributes in public profiles, shows a correlation between the amount of information revealed in social network profiles, specific occupations or job titles and use of pseudonyms. It is possible to identify certain patterns regarding how and when users reveal precise information~\cite{chen-et-al}. Finally, aggregating this information can lead an attacker to obtain direct contact information by cross-linking the obtained features with other publicly available sources, such, for example, online phone directories.\\
\\
\noindent
A famous method for information correlation was presented by Alessandro Acquisti and Ralph Gross~\cite{acquisti-et-al}. Leveraging on the correlation between individuals' Social Security numbers and their birth data, they were able to infer people Social Security numbers by using only publicly available information.\\
\\
\noindent
Privacy attackers can also exploit loose privacy settings of a user's online social connections, taking advantage of how humans interpret messages and interact with one another~\cite{cryto-gram}, developing semantic attacks~\cite{Kumaragur-et-all}.\\
Therefore, mechanism helping to promote coordinated privacy policies could be more efficient to count attacks~\cite{brown-et-all}. \\
\\
\noindent
Accurate coordinated policy could also warn users of which third party application they authorise to access their data. Social networking platforms, in fact, expose users' privacy to possible attacks by allowing third party application that access their data to be able to replicate it.\\ 
Sandboxing techniques could be implemented allowing users to share information among social relationships, while also helping third party application to securely aggregate data according to differential privacy properties~\cite{viswanath2012keeping}. \\
\\
\noindent
Users should be allowed to choose an appropriate level of privacy for their needs and should be made aware of unwanted access to their data. This would permit protection of personal information that is being collected by mobile devices, including the derived inferences that could be drawn from the data. Semantic Web technologies can be implemented to specify high-level, declarative policies describing user information sharing preferences~\cite{jagtap2011preserving}. \\
\\
\noindent
Users, in fact, consider three main deciding factors when consulted about how and to what extent they are willing to disclose personal and sensitive information, especially information about their location, to social relations~\cite{consolvo2005location}. These factors were: who was requesting a particular information, why that information was requested, and what level of detail would be most useful to the requester. \\
\\
\noindent
This aspect of users' perception of sensitive information disclosure is particularly relevant when it has been shown~\cite{toch2013locality} that knowing a user location is used as a grounding mechanism in applications that lets users interact with their nearby. Geo-tagged information set the basis for a platform for honest and truthful signals in the process of forming new social relations.\\
At the same time, geo-localised information attached to users' activities can be used, by an attacker, to derive models of user mobility and provide data for context-­aware applications and recommendation systems~\cite{melia-segui-et-al}. This information can also be used to cluster communities with different preferences and interests into different geographical communities~\cite{zhu-et-al}. \\
\\
Also, while some social networking applications use some form of obfuscation of the users' actual positions, precise location information can be still be derived. An attacker 
could use the partial information to identify a user's real position even when their exact coordinates are hidden or obfuscated  by  various  location  hiding  techniques~\cite{li2014all}.
\\
\noindent
Therefore, we consider the problem of identifying the geographical position of a node in a network given their imprecise relative distance. This particular problem has being studied extensively in the literature of wireless and sensors networks. \\
\\
\noindent
Geometric relations among  distances  between  nodes  can be formulated as  equality  constraints~\cite{cao2005localization}. This approach is used to study the localisation problem with imprecise distance information in  sensor networks. The error in the inaccurate distances between sensor nodes and anchor nodes starting is estimated from a set of algebraic constraints. The problem is formulated as a least squares problem with the objective to  minimise  the  sum  of  the  squared  errors. Other  objective functions  are  also adoptable  depending on the specific application context. \\
\\
\noindent
Estimating the location of a node is also possible in situation where their distance, or distance measurement, is known with a relatively poor precision. A node of a randomly placed wireless sensor network can, in fact, be accurately track and targeted even with poor distance measurement accuracy~\cite{nagpal2003organizing}.\\
\\
\noindent
An attacker could also compromise a number of nodes in a network to verify, or disclose, the position of a particular node. 
In this particular scenario, n entity, the prover, claims and wishes to prove its location. On the other side a different entity, the verifier, has the role of verifying such location claim. The verifier measure the distance to the prover and can approve or not the claim that the prover must be must be within a circle of a certain radius. By using an arbitrary number of verifier at the same time, the actual location of the prover is determined~\cite{chiang2009secure}. 

\section{Classifying privacy violations}

We follow the approach of Daniel J. Solove in ~\cite{solove2006taxonomy} to classify privacy violations in four main categories (Table ~\ref{tableviolations}). These are: 

\begin{enumerate}
 \item Information collection
\\ 
 \item Information processing
 \\
 \item Information dissemination
 \\
 \item Invasion
 \end{enumerate}
 
 \subsection{Information collection}
 Information collection results from activities such as surveillance, interrogation or information probing. It refers to actions aimed at watching, reading, listening, recording of individual activities or data about activities. It also refers to direct questioning of individuals or inference of information from data about them.
 
 \subsection{Information processing}
 Information processing concerns the aggregation and identification of data. Failure to provide data security and the possibility for users to know who has accessed their data. This also includes secondary use of data to which the user has not been informed.
 
 \subsection{Information dissemination}
 Information dissemination includes activity such as breach of confidentiality, unwanted disclosure and exposure of information. This also includes increased accessibility to individual's information, appropriation and distortion of data about people. Information dissemination defines the very action of breaking the promise of keeping information confidential. It therefore implies actions aimed at the revelation of information about an individual that can change the image of that person within a group, including appropriation of identity information and dissemination of false or misleading facts. 
 
 \subsection{Invasion}
 Invasion is the threat of intrusion of an entity into someone's private life and it includes acts that are said to disturb one's tranquillity or solitude.

\begin{table*}[t]
\centering
\caption{Classification of privacy violations}
\def\arraystretch{2.0}
\begin{tabulary}{\linewidth}{ | p{1.5cm} | p{5cm}  | p{10.5cm} | }
  \hline
   Violation				& Activities																	&	Actions  																						\\
  \hline
  \hline
Collection				& Surveillance; Information probing; Interrogation										&	 Watching, listening, recording of individuals activities. Questioning individual directly. Inferring information from data.	\\ 
Processing 			& Aggregation; Identification; Insecurity; Secondary use; Exclusion;							&       Gathering of data about individuals. Identification of physical identities from online data. Carelessness in protecting data.  Failure in allowing users to know who has accessed to their data.	 \\  
Dissemination 			& Breach of confidentiality; Disclosure; Exposure; Increased accessibility; Appropriation; Distortion;	&       Breaking the promise of keeping the information confidential. Revelation of information about an individual that impacts the way other see them. Appropriation of identity information. Dissemination of false or misleading information. Transfer of personal data to third party or threat to do so. \\
Invasion		 		& Intrusion of someone's private life													&       Acts that can disturb one's tranquillity or solitude. \\
  \hline 
\end{tabulary}
\label{tableviolations}
\\[2.5pt]
 \begin{flushleft}
The table summarises the classification used to categorise privacy violation in proximity-based social application.
\end{flushleft}
\end{table*}

\section{Modelling the location probe method}

\begin{figure}[t]
\centering
\includegraphics[width=90mm]{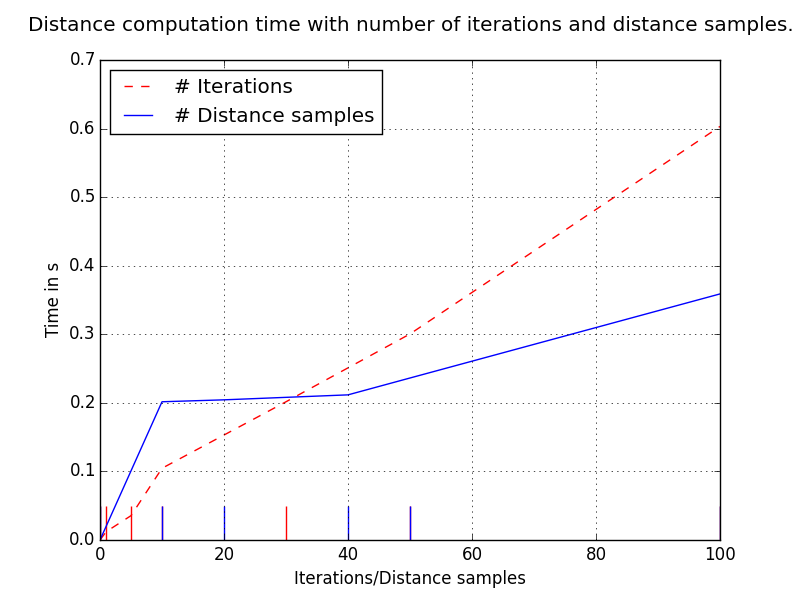}
\caption{The image illustrates the time needed to compute a user position estimation based on the number of distance samples and the number of iterations of the algorithm. It is important to note how the number of distance samples does not affect the algorithm performances. The example was executed on a Apple Computer with 3 GHz Intel Core i7 Processor.
\label{comp_time}}
\end{figure}

Proximity based social application collect users' positions and share their relative distances. We show how it is possible to build a multilateration attack able to identify the actual user position with arbitrary precision.\\
\noindent
Multilateration is a navigation technique, often used in radio navigation systems, based on the measurement of the difference in distance to two or more stations, whose locations are known. The stations also produce a certain signal at a known time.\\
\noindent
In our scenario, the signal is replaced by the user distance from the attacker and time is given by the timestamp of the user latest activity. \\
\noindent
Please note that, multilateration is not concerned with measurements of absolute distance or angle between parties, but with measuring the difference in distance between two stations which results in an infinite number of locations that satisfy the measurement. All these possible locations form a hyperbolic curve. Multilateration therefore relies on multiple measurements to locate the exact location along that curve. In fact, a second measurement taken to a different pair of stations will produce a second curve, which intersects with the first. When the two curves are compared, a small number of possible locations are revealed.\\
\\
\noindent
If the attacher is able to retrieve an arbitrary number of samples of the user distance, either by changing their relative location or by sampling their distance with the victim with a number of malicious mobile client infiltrating the platform, the multilateration attack can be made arbitrary precise.\\
\\
\noindent
Our location probe method uses a simple multilateration algorithm. At the first step, locations expressed as longitude and latitude coordinates are translated to cartesian coordinates. We then calculate the estimated distance and minimise the linear norm between calculated distance and estimated distance by sensing the total error. We could have considered the total squared error between the estimated and actual distance, however in this paper we have concentrated on demonstrating that the attack is actually feasible, rather then on accuracy or performance of the algorithm (Fig. ~\ref{comp_time}).

\section{Modelling the user activity profile}

We model the user's activity as series of events belonging to a certain identity. Each event is a document containing different information. We can formally defined this an hypermedia document i.e. an object possibly containing graphics, audio, video, plain text and hyperlinks. We call the hyperlinks selectors and we use these to build the connections between the user's different identities or events. Each identity is a profile that the user has created onto a service or platform. This can be an application account or a social network account, such as their LinkedIn or Facebook unique IDs.\\
\\
\noindent
Each event is the result of the user performing an action. For the purpose of this study we have consider an action as resulting using an application or a service. An action is the activity of interacting with a mobile application or \emph{liking} a resource on a social network, i.e. directly expressing an interest, or the fact that a user has updated their location at a certain time.\\
\noindent
Formally it is possible to model the graph of the events pertaining to a user as an hypergraph, where each edge can connect any number of vertices, and the root is first event in the series.  A hypergraph $H$ is a pair $H = (X,E)$ where $X$ is a set of nodes (the events in the model), and $E$ is a set of non-­empty subsets of $X$ called hyperedges or edges. Hypergraphs are a generalisation of a graph structure and provide a reasonable representation of the connections between the different events resulting of the actions performed by the user.\\
\\
\noindent
We find that this model is able to express the user's online footprint as a collection of traces left across different services. Furthermore by using a hypergraph model we are able to grasp the connections between the different profiles and features. \\
\noindent
This results in the possibility to profile users based on chosen selectors. For example we might want to trace all users who have been in the radius of 500 meters to a certain location, or all the users in a certain neighbourhood who \emph{like} a selected Facebook page.

\subsection{Adversary model}

In view of the assumptions described in the previous section, our privacy attacker boils down to an entity that aims to identify users and link their online profile to their physical identity. To achieve this objective the attacker posses a Facebook profile. This profile is used in the first place to register to the application analysed in this study since all three use Facebook Login as a personalised way for user to register and sign in. 

\section{Experimental results}

We have analysed 250 users from a set of social proximity applications (Table ~\ref{tableapps}). All applications examined are matchmaking mobile platforms which uses geolocation technology. Users can use their location and preferences to search for interesting people in a specific radius. All applications use Facebook profiles to allow their users to login but also to gather basic information and analyse users' social graph. The information collected are then used to match candidates who are most likely to be compatible based on geographical location, number of mutual friends, and common interests. 

\begin{table*}[t]
\centering
\caption{Information regarding the applications analysed}
\def\arraystretch{2.0}
\begin{tabular}{| c | c | c | c | c | c | c | c | c |}
  \hline
    Application 				& Users 							& 	 Facebook ID 		& 	 Location	 	& 	 Distance		&  	User Pref.      & 	 Full Name 	&  	Birthdate 	     & Allow user tracking \\
  \hline
  \hline
    Tinder~\cite{tinder}		& 10 Million active~\cite{tinderusers} 	& 	\ding{55} (1) 		& 	\ding{55} 		 &	 \ding{51} 		& \ding{51}	      & 	\ding{55} (2)	&	\ding{55} (3)  &		\ding{51}		\\ 
    Happn~\cite{happn} 		& ~ 700.000 ~\cite{happnusers} 		& 	\ding{51} (1)		& 	\ding{55} 	 	 &	 \ding{51} 		& \ding{51}	      & 	\ding{55} (2)	&	\ding{55}	     &		\ding{51}		\\  
    Lovoo~\cite{lovoo} 		& ~ 24 Million registered~\cite{lovoousers}& 	\ding{55} (1)		& 	\ding{55}		 & 	\ding{51} 		& \ding{51}	      & 	\ding{55} (2)	&	\ding{55}	     &		\ding{51}		\\   
    Grinder~\cite{grindr} 		& ~ 2,35 Million active~\cite{grindrusers}	& 	\ding{55} (1)		& 	\ding{55}	 	 & 	\ding{55} 		& \ding{51}	      & 	\ding{55}		&	\ding{55}	     &		\ding{55} 		\\  
    Badoo~\cite{badoo}	 	& 200 million registered~\cite{badoousers}& 	\ding{55} (1)		& 	\ding{55}	 	 & 	\ding{51}(4) 	& \ding{51}	      & 	\ding{55} (2)	&	\ding{55}	     &		\ding{51} (6)	\\  
  \hline 
\end{tabular}
\label{tableapps}
\\[2.5pt]
 \begin{flushleft}
(1) Facebook ID is not exposed directly but it can be identified by crossing information like the user Facebook's likes, first name and year of birth. \\
(2) Only first name is shared. \\
(3) A fuzzy birthdate randomised in a range of two weeks is used. Real birthdate can be inferred by using Facebook Graph Search, depending on the victim's Facebook privacy settings. \\
(4) Offers option not to share distance. \\
(5) Asks for zodiac sign. \\
(6) Distance is shared for some users so it is theoretically possible.\\
\end{flushleft}
\end{table*}

These applications present the user with the possibility to interact with other users by starting conversation or expressing their interests in them. 

\subsection{Information collection}
Information collection is possible on these applications through different techniques. For the purpose of this study we have intercepted APIs call from mobile devices through Men In The Middle (MITM) attack in some occasions, and interacted with the APIs directly in other occasions.
\noindent
It is important to note that even when the application prevents an attacker from exploiting their APIs, a malicious entity could still use a multitude of profiles to cross gather information about users on the platforms.

\subsection{Information processing}
We have performed two types of attack on the set of application examined, namely a multilateration attack and a social graph attack.

\subsubsection{Multilateration attack}
Once we posses the user's id on the specific application we are able to query their APIs and update our information about the user constantly. Furthermore we are also able to change our own location on the platform to a certain extent. 
\noindent
By measuring the relative distance to the victim we were able to identify their actual position with arbitrary precision. Furthermore, the same technique was used to ~\emph{follow} users across a specific amount of time by retrieving their profile information at regular interval.
\noindent
This type of attacks can be easily overlooked in densely populated cities but might become a serious privacy breach in rural areas.

\begin{figure}[t]
\centering
\includegraphics[width=90mm]{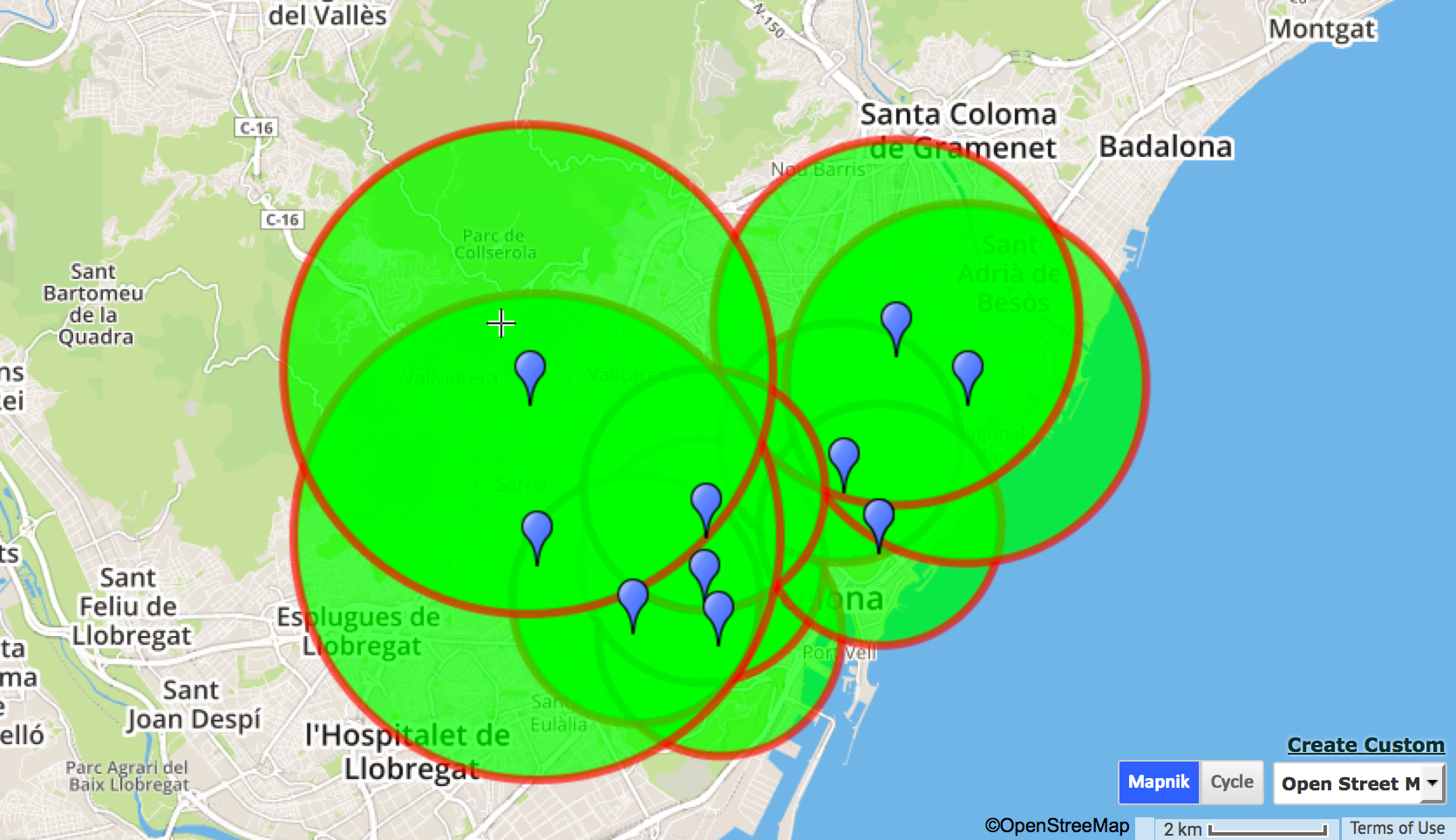}
\caption{The image illustrates location samples with radii used to compute actual position estimation for one user across the city of Barcelona, Spain.
\label{mlat}}
\end{figure}

\subsubsection{Hyper graph attack}

\begin{figure}[t]
\centering
\includegraphics[width=90mm]{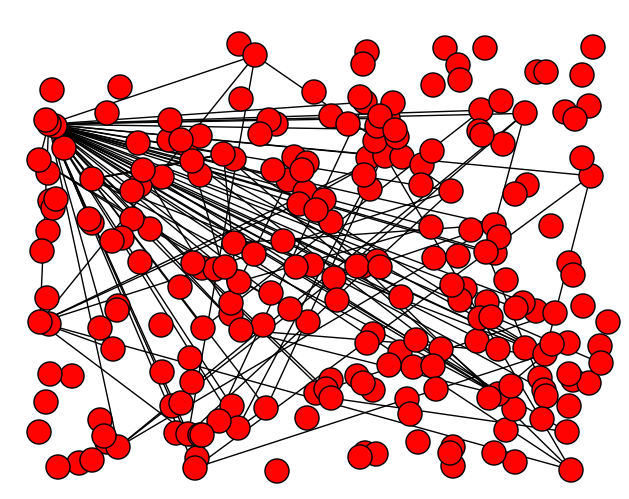}
\caption{The image shows how it is possible to show connections for the population of users on Tinder for a certain area. Here we have collected Facebook pages liked by users in Barcelona and connected users or group of users, if they like the same page.\label{connections}}
\end{figure}

\begin{figure}[tb!]
\centering\hspace*{\fill}
{\includegraphics[width=90mm]{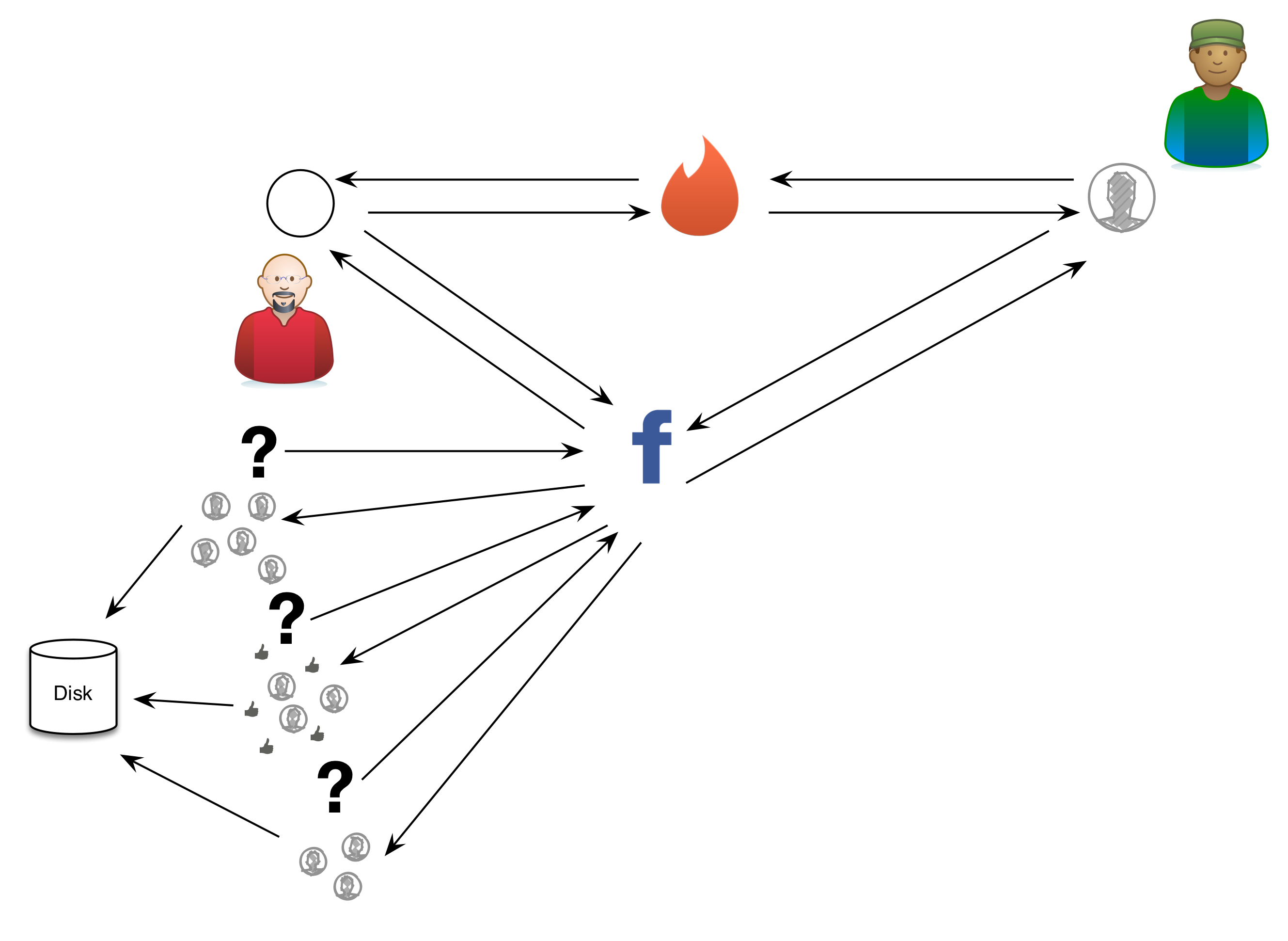}%
\label{graph}}\hspace*{\fill}
\caption{The image represents a Social Graph attack where an attacker sends queries Facebook asking question about a Tinder profile. The attacker is able to restrict the pool of potential candidates and eventually identify the victim's actual Facebook id. Furthermore the attacker is able to store information about the user that can be updated at a later time by querying the third party application.}
\label{FGS}
\end{figure}

The application examined for the scope of this study use the user's Facebook token to authenticate and/or authorise the application to request and obtain certain information about the user. 
An attacker could then use their own Facebook profile token to make request to the application server through their APIs, pretending to send the request from the app installed in a mobile device. This allows the attacker to receive all the information that users have shared with the platform and that are constantly exchanged with the application. \\
\\
\noindent
When the victim's Facebook id is shared through the application, the attacker can directly access and potentially use information publicly shared through the Facebook profile. In this situation the attacker could easily construct a complete graph of the user's preferences and social connections through the information that are public available through Facebook APIs.\\
\\
\noindent
When the victim's Facebook id is not directly shared, the application still disclose some information about the victim. This information include: the user first name and a set of photos, birthdate, randomised in a range of 15 days, and the Facebook pages that both the victim and the attacker have liked. \\
\\
\noindent
The victim preferences could then be used to identifies their Facebook profile. It is in fact estimated that Facebook posses 1.35 billion active users, of these, between 10\% and 7\% like one of the top 10 Facebook pages with most likes~\cite{pagedata}. We have collected a set of 250 Tinder users only in the city of Barcelona, of these 20\% where sharing at least one interest with the attacker profile (Fig. ~\ref{connections}).\\
\\
\noindent
Furthermore Facebook graph search allows any users to answer certain information about Facebook profiles. An example of a graph search on Facebook could be: \emph{People who like Shakira and are named "John" and like Manchester United and been born in 1979}. This will create a pool of potential candidates. The list can be reduced by using Facebook reverse graph search, i.e. search for \emph{Interests liked by people who like Shakira and are named "John" and like Manchester United and been born in 1979}.  This will instead return a list of interests that the attacker can like on Facebook. Therefore the attacker will return to query Tinder and find out if the number of interests in common with the victim has grown and which pages they now have in common. The attacker can therefore use the new information to further identify the victim profile on Facebook and potentially their friends (Fig. ~\ref{FGS}).\\
\\
\noindent
It is important to note that some applications might request information outside of Facebook public profile. Therefore even if the victim has tailored their privacy settings to prevent some information to be leaked, the application can be used to access data that would be otherwise be kept private.

\subsection{Information dissemination}

Proximity-based social applications, in their current implementation, represent a gateway to access data about individuals. Information dissemination can therefore be accomplished both for large group of people with the purpose of targeting them, as well as for specific victims. 
Identifying and disclosing the presence of a certain person on a match making application could be enough to influence the opinion of that individual among their social relationships. 

\subsection{Invasion}
Once a user location has being inferred, we can continue tracking the same users and their preferences for an unlimited amount of fetches. This could easily lead to identification of the user habit and whereabout at different moment of the day, possibly uncovering their home and work locations and more information about the user. 

\section{Mitigation possibilities}

Application developers could implement a number of techniques that would mitigate the actions of a possible attacker.\\
\noindent
Firstly, in their current implementation, the applications examined probe the user device for location information with the maximum precision possible. This information is then transferred to the server and the relative distance between users is returned to be displayed. Yet, for most of the application functionalities this precision is not needed, and a lower precision could be used and sent to the server. This would make position attacks more difficult to perform. \\
\noindent
Secondly, to sparkle interest between users, social proximity application often share common Facebook pages between parties involved. This information can then be used to easily identify unique Facebook accounts. Instead, the app could opt to display only the category of interest to which the Facebook page belongs. This way a possible attacker would not know what actual pages the user has liked.\\
\noindent
Thirdly, an individual birthdate if combined with their location and first and/or last name can be used to infer sensitive information about them. Therefore even sharing the user's zodiac sign with passive observer need to be considered potentially dangerous for the final user's privacy.

\section{Conclusion}

A new class of social application uses the users' actual location to provide personalised recommendation and allow for new interactions especially in urban settings. We have shown how these applications can expose their users to different privacy attacks that can be easily overlooked.\\
\noindent
We have analysed a set of  popular dating application, and observed how proximity-based social applications have access to certain identity information that could lead a possible privacy attacker to easily identify users on Facebook and link their online profiles to physical identities. \\
\\
\noindent
Furthermore we have shown how users constantly sharing their relative distance to other users can be ~\emph{followed} by an attacker in their movement without their knowledge. We have demonstrated how this information can be used for a multilateration attack with arbitrary precision. There is in fact no restriction to the number of distance samples that a possible attacker might be able to measure.\\
\noindent
We followed a formal framework to identify the classes of privacy violation to which users are subjected to without being aware of it and we have shown how these violations can all be carried out for the applications examined.\\
\\
\noindent
This shows how using third party profiles to provide access to a specific applications may cause a security \emph{honey pot} for a possible attacker. \\
\\
\noindent
We have also stressed how In order to make the registration process easier, these applications often leverage on third party services to provide a login mechanism, while at the same time acquiring certain private information about their new users. The third parties used are often services such as Facebook or Google, and the information accessed concern the public profile of the users on such platforms. \\
\noindent
While this technique certainly allows people to quickly sign up to an application and create a new profile, it also creates different privacy threats for users of such services. Primarily, it concerns who can gain access to such data and how information shared with third parties can also be stored and eventually transferred without the user explicit consent.\\
\noindent
We have then used Facebook graph search to build a hyper graph of the user identity starting from few information that were shared through a third application. This shows how each information can be used as a selector to further identify a different piece of the whole user identity and can be used to target the user in real life.\\

\section*{Acknowledgment}
This work was partly supported by the Spanish Government through projects CONSEQUENCE (TEC2010-20572-C02-02) and EMRISCO (TEC2013-47665-C4-1-R).

\bibliographystyle{IEEEtran}

\bibliography{Bibliography/StringAbbreviated,Bibliography/Security,Bibliography/InfoTheory,Bibliography/LosslessCoding,Bibliography/LossyCoding,Bibliography/MathStatSigPro,Bibliography/Classification,Bibliography/Applications,Bibliography/ReferencesTRIPP,Bibliography/SemanticWeb,Bibliography/InsubriaReferences,Bibliography/rfc,Bibliography/Silvia_bibliography,Bibliography/thesis_proposal.bib}

\end{document}